# Chiral and Steric Effects in Ethane: A Next Generation QTAIM Interpretation

Zi Li, Tianlv Xu[1], Herbert Früchtl[2], Tanja van Mourik[2], Steven R. Kirk[*1] and Samantha Jenkins[*1]

[1]*Key Laboratory of Chemical Biology and Traditional Chinese Medicine Research and Key Laboratory of Resource National and Local Joint Engineering Laboratory for New Petro-chemical Materials and Fine Utilization of Resources, College of Chemistry and Chemical Engineering, Hunan Normal University, Changsha, Hunan 410081, China*

[2]*EaStCHEM School of Chemistry, University of Saint Andrews, North Haugh, St Andrews, Fife KY16 9ST, Scotland, United Kingdom.*

email: steven.kirk@cantab.net
email: samanthajsuman@gmail.com

We introduce a development of next generation quantum theory of atoms in molecules (NG-QTAIM) for an investigation of the chirality of ethane and discover a $\mathbf{Q}_\sigma$ isomer in addition to $\mathbf{S}_\sigma$ and $\mathbf{R}_\sigma$ stereoisomers in the stress tensor trajectory $\mathbb{U}_\sigma$-space. The $\mathbf{Q}_\sigma$ isomer is defined to be a 'null-isomer' since the value of the chirality-helicity function ≈ 0. The presence of chiral contributions suggests that steric effects, rather than hyper-conjugation, explain the staggered geometry of ethane. The steric effects, within the NG-QTAIM interpretation, are reduced by a factor of two using an applied electric-field directed down a C-H bond.

# 1. Introduction

Recently, some of the current authors used next generation QTAIM (NG-QTAIM) to quantify a chirality-helicity measure[1], which is an association between molecular chirality and helical characteristics known as the chirality-helicity equivalence by Wang[2], consistent with photoexcitation circular dichroism experiments. Wang stated that the origin of this helical character was not provided solely by molecular geometries or attributable to steric hindrance alone, but would require insight provided from the electronic structure. Recently, the interdependence of steric-electronic factors was discovered to be more complex[3] than was discernable from the molecular structures for the helical electronic transitions of spiroconjugated molecules[4,5].

Recent experiments by Beaulieu *et al.* that utilized coherent helical motion of bound electrons on neutral molecules demonstrated the need for a better understanding of the behavior of the charge density redistribution [6]. Banerjee-Ghosh *et al.* also demonstrated that charge density redistribution in chiral molecules, rather than spatial effects[7], is responsible for an enantiospecific preference in electron spin orientation, consistent with Wang.

If steric effects are generally implicated in explanations of chirality, albeit with limitations as indicated by Wang, it seems reasonable to use an electronic charge density based analysis to investigate steric effects. A recent investigation to quantify the chirality-helicity equivalence found that what is generally understood as chirality is only *part* of the complete understanding of the term 'chirality' when referring to conventionally chiral molecules. Instead of only being concerned with the asymmetry of the chiral carbon, which is generally referred to as chirality, we demonstrated that the bond-axiality, which is concerned with the motion of the bond critical point (*BCP*) along the torsion bond, should also be considered[5]. We suggest therefore that the chirality-helicity equivalence can be used to understand the staggered conformation of ethane. The staggered conformation of the relaxed ethane structure is still controversial, with energy-based analysis providing contrasting explanations for the cause of the staggered conformation. The most commonly accepted reason is that of steric (repulsion) effects[8–17]. Pophristic and Goodman[18] however, argued that hyper-conjugation rather than steric repulsion leads to the staggered structure of ethane. A point of emphasis of their analysis was that eliminating repulsive interactions was a factor in determining the staggered ethane structure and was undertaken by the inclusion of skeletal expansion, i.e. explicit C–C bond lengthening, which means that the C-C torsion does not comprise pure rotation.

Recently, Shaik *et al.* considered an electric-field (**E**-field) as a 'smart reagent' in a range of reactions, for the control of reactivity and structure for chemical catalysis[19]. We recently considered formally achiral glycine subjected to an **E**-field using a vector-based interpretation of the total electronic charge density distribution[20]. This work on glycine followed on from the earlier investigations on glycine by Wolk *et al.* on the application of an ±**E**-field to the alpha carbon atom (C1), which induced symmetry-breaking changes to the length of the C-H bonds [21]. Our NG-QTAIM investigation demonstrated the possibility to control the magnitude of the chirality induced by the application of the ±**E**-field.

In this investigation we will test the effect of the absence and presence of an **E**-field directed along each of the three C-H bonds attached to the 'chiral' carbon (C1) of ethane on the NG-QTAIM properties. The application of the **E**-field, separately along the three C-H bonds, does not provide sufficient symmetry breaking to create structural stereoisomers despite inducing changes to the lengths of the C-H bonds, see **Scheme 1**. We will use the **E**-fields: $\pm 50 \times 10^{-4}$ a.u. and $\pm 100 \times 10^{-4}$ a.u, which are easily accessible experimentally, for example within a Scanning Tunneling Microscope (STM). In this investigation we will consider the *non-energy-based* NG-QTAIM alternative to steric effects and hyper-conjugation of ethane to gain insights into the staggered structure of ethane by locating any $\mathbb{U}_\sigma$-space isomers and or stereoisomers.

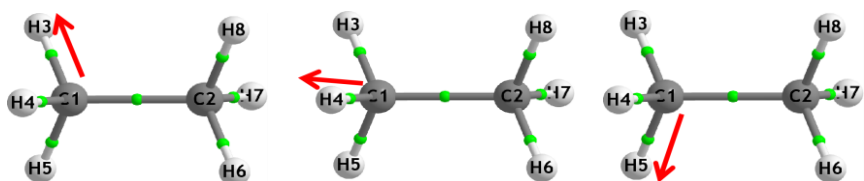

**Scheme 1.** The molecular graphs of ethane with arrows indicating the directions of the positive electric (+)**E**-field of the C1-H3 *BCP* bond-path (left panel), C1-H4 *BCP* bond-path (middle panel) and C1-H5 *BCP* bond-path (right panel). The green spheres indicate the bond critical points (*BCP*s).

## 2. Theoretical Background and Computational Details

The background of QTAIM and next generation QTAIM (NG-QTAIM)[22–28] is provided in the **Supplementary Materials S1**, including the procedure to generate the stress tensor trajectories $\mathbb{T}_\sigma(s)$.

Bader's formulation of the stress tensor[29] and NG-QTAIM is a standard option in the AIMAll QTAIM package [30] and is used in this investigation because of the superior performance of the stress tensor compared with the Hessian of $\rho(\mathbf{r})$ for distinguishing the S and R stereoisomers of lactic acid[31].

The chirality $\mathbb{C}_\sigma$ is defined as the difference in the maximum projections (the dot product of the stress tensor $\mathbf{e}_{1\sigma}$ eigenvector and the *BCP* displacement **dr**) of the $\mathbb{T}_\sigma(s)$ values between the counter-clockwise (CCW) and clockwise (CW) torsion θ: $\mathbb{C}_\sigma$ = [(**e**$_{1\sigma}$·**dr**)$_{max}$]$_{CCW}$ - [(**e**$_{1\sigma}$·**dr**)$_{max}$]$_{CW}$. The chirality $\mathbb{C}_\sigma$ quantifies the bond torsion direction CCW vs. CW, i.e. *circular* displacement, where the largest magnitude stress tensor eigenvalue ($\lambda_{1\sigma}$) is associated with **e**$_{1\sigma}$. The **e**$_{1\sigma}$ corresponds to the direction in which the electrons at the *BCP* are subject to the most compressive forces and therefore will be the direction along which the *BCP* electrons will be most displaced when the *BCP* is subjected to torsion[32]. The eigenstructures of the stress tensor **σ(r)** and QTAIM Hessian of $\rho(\mathbf{r})$ rarely coincide as can be seen in the {*q$_\sigma$*,*q$_\sigma$*'} and {*q*,*q*'} path-packets respectively, which are constructed from the corresponding eigenvectors and eigenvalues, see the **Supplementary Materials S4**. In the $\mathbb{U}_\sigma$-space distortion set {$\mathbb{C}_\sigma$,$\mathbb{F}_\sigma$,$\mathbb{A}_\sigma$} the bond-flexing $\mathbb{F}_\sigma$, defined as $\mathbb{F}_\sigma$ = [(**e**$_{2\sigma}$·**dr**)$_{max}$]$_{CCW}$ - [(**e**$_{2\sigma}$·**dr**)$_{max}$]$_{CW}$, provides a measure of the 'flexing-strain' that a bond-path is under when subjected to an external force such as an **E**-field. The bond-axiality $\mathbb{A}_\sigma$, which provides a measure of the chiral asymmetry, is defined as $\mathbb{A}_\sigma = [(\mathbf{e}_{3\sigma}\cdot\mathbf{dr})_{max}]_{CCW} - [(\mathbf{e}_{3\sigma}\cdot\mathbf{dr})_{max}]_{CW}$. This quantifies the direction of *axial* displacement of the bond critical point (*BCP*) in response to the bond torsion (CCW vs. CW), i.e. the sliding of the *BCP* along the bond-path[33]. The signs of the chirality $\mathbb{C}_\sigma$, bond-flexing $\mathbb{F}_\sigma$ and bond-axiality $\mathbb{A}_\sigma$ determine the dominance of **S**$_\sigma$ ($\mathbb{C}_\sigma > 0$, $\mathbb{F}_\sigma > 0$, $\mathbb{A}_\sigma > 0$) or **R**$_\sigma$ ($\mathbb{C}_\sigma < 0$, $\mathbb{F}_\sigma < 0$, $\mathbb{A}_\sigma < 0$) character, see **Table 1**. The intermediate results for the $\mathbb{U}_\sigma$-space distortion sets $\{\mathbb{C}_\sigma,\mathbb{F}_\sigma,\mathbb{A}_\sigma\}$ are provided in the **Supplementary Materials S5**.

An additional null-chirality assignment **Q**$_\sigma$ ($\approx 0$ chiral character) occurs where a plot of ellipticity ε vs torsion θ displays CCW and CW portions that are symmetrical about torsion θ = 0.0°, see **Figure 1** (left-panel). The choice of ± sign is therefore not used with the chirality assignment **Q**$_\sigma$ as it is for the **S**$_\sigma$ and **R**$_\sigma$ assignments, where for the latter mirror symmetry is only present for the **S**$_\sigma$ CCW vs **R**$_\sigma$ CW and **S**$_\sigma$ CW vs **R**$_\sigma$ CCW plots of ellipticity ε vs torsion θ, see **Figure 1** (right-panel).

The chirality-helicity function $\mathbb{C}_{helicity} = \mathbb{C}_\sigma|\mathbb{A}_\sigma|$, i.e. the simple numerical product of the chirality and the magnitude of the bond-axiality $\mathbb{A}_\sigma$, can be used to determine the nature of any chiral behaviors present in ethane in the absence or presence of an **E**-field. The tabulated form of the $\mathbb{C}_{helicity}$ is expressed as $[\mathbb{C}_{\sigma T},\mathbb{A}_\sigma]$, where total chirality $\mathbb{C}_{\sigma T}$ may possess **Q**$_\sigma$, **S**$_\sigma$ or **R**$_\sigma$ chirality assignments and the bond-axiality $\mathbb{A}_\sigma$ may possess **S**$_\sigma$ or **R**$_\sigma$ chirality assignments.

Low/high values of the chirality $\mathbb{C}_\sigma$ for the C1-C2 *BCP* are associated with low/high steric effects due to the absence/presence of an asymmetry in the CCW vs CW motion along the most preferred direction of $\rho(\mathbf{r_b})$ accumulation (**e**$_{1\sigma}$). For instance, an artificially imposed eclipsed ethane geometry may correspond to values of an overall $\mathbb{C}_\sigma$ ($\approx 0$) because of the equal preference for torsion of CCW vs. CW motion. We associate low/high values of the bond-axiality $\mathbb{A}_\sigma$ for the C1-C2 *BCP* bond-path with a low/high hyper-conjugation contribution due to the absence/presence of an asymmetry in the CCW vs CW bond-path stretching. The NG-QTAIM interpretation of ethane as an achiral molecule requires calculation of all the symmetry inequivalent stress tensor trajectories $\mathbb{T}_\sigma(s)$ through the torsion C1-C2 *BCP*, which results in a complete set of ethane $\mathbb{U}_\sigma$-space isomers, with possible chirality assignments **Q**$_\sigma$, **S**$_\sigma$ or **R**$_\sigma$.

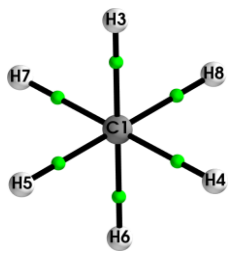

**Scheme 2**. An axial view down the torsion C1-C2 *BCP* bond-path of the molecular graph of ethane with the atomic numbering scheme used for the dihedral angles in the construction of the torsion C1-C2 *BCP* stress tensor trajectories $\mathbb{T}_\sigma(s)$. The H3, H4 and H5 atoms are bonded to the C1 atom and the H6, H7 and H8 atoms are bonded to the C2 atom.

In this investigation we will include all contributions to the $\mathbb{U}_\sigma$-space chirality by considering the entire bonding environment of the 'chiral' carbon atom (C1) by constructing all nine torsion C1-C2 *BCP* $\mathbb{T}_\sigma(s)$ using dihedral angles that include the C1 atom, see **Scheme 2** and the Computational Details section. The linear sum of the individual components of the symmetry inequivalent $\mathbb{U}_\sigma$-space distortion sets $\sum\{\mathbb{C}_\sigma, \mathbb{F}_\sigma, \mathbb{A}_\sigma\}$ will be calculated to provide the resultant chiral character for the ethane molecular graph. Then $\mathbb{C}_{helicity}$ (= $\mathbb{C}_\sigma|\mathbb{A}_\sigma|$) of the resultant will be calculated as the linear sum $\sum\{\mathbb{C}_\sigma, \mathbb{F}_\sigma, \mathbb{A}_\sigma\}$ as well as the sum of the individual $\mathbb{C}_{helicity}$; values of $\mathbb{C}_{helicity} \approx 0$ correspond to the null-chirality assignment $\mathbf{Q}_\sigma$, see **Table 1**.

*Computational Details*

The ethane molecular structure was initially geometry-optimized with 'verytight' convergence criteria at the B3LYP/cc-pVQZ level of DFT theory using Gaussian 09.E01 [34] with an 'ultrafine' integration grid. The wavefunctions were converged to < $10^{-10}$ RMS change in the density matrix and < $10^{-8}$ maximum change in the density matrix. All subsequent **E**-field optimization, torsion and single-point steps used identical convergence criteria. An iterative process is used to create the **E**-field induced isomers; this is undertaken by directing an **E**-field parallel (+**E**-field) or anti-parallel (-**E**-field) to each of the C1-H3 *BCP*, C1-H4 *BCP* or C1-H5 *BCP* bond-paths, see **Scheme 1**. The label $A_{3\mathbf{E}\sigma}$ is assigned where the **E**-field is applied to the C1-H3 *BCP* bond-path, the label $A_{4\mathbf{E}\sigma}$ for an **E**-field applied to the C1-H4 *BCP* bond-path length and the label $A_{5\mathbf{E}\sigma}$ for an **E**-field applied to the C1-H5 *BCP* bond-path. Each of the $A_{3\mathbf{E}\sigma}$, $A_{4\mathbf{E}\sigma}$ and $A_{5\mathbf{E}\sigma}$ isomers are subjected to a two-step iterative process consisting of (a) a molecule alignment step: the alpha C1 atom is fixed at the origin of the coordinate frame, whereas the selected C-H bond is aligned along a reference axis with the positive direction of the axis from C to H and the C atom consistently aligned in the same plane. This is followed by (b): a constrained optimization step with the selected **E**-field applied along the reference axis. The default G09 sign convention for the field relative to the reference axis is used. This two-step process is repeated ten times, to ensure consistency of the **E**-field application direction and the chosen bond (C1-H3, C1-H4 or C1-H5) direction. The resulting molecular structures are then used in the subsequent torsion calculations where the C1-H3, C1-H4 and C1-H5 bond lengths are constrained to their **E**-field-optimized values. The ethane is subjected to **E**-fields = $\pm50\times10^{-4}$ a.u. and $\pm100\times10^{-4}$ a.u. before the molecular structure undergoes a torsion to construct the trajectories $\mathbb{T}_\sigma(s)$ from the series of rotational isomers -180.0º $\leq \theta \leq$ +180.0º for the torsional C1-C2 *BCP* of ethane. We determine the direction of torsion as CCW (0.0º $\leq \theta \leq$ +180.0º) or CW (-180.0º $\leq \theta \leq$ 0.0º) from an increase or a decrease in the dihedral angle, respectively. The $\mathbb{T}_\sigma(s)$ for all nine possible ordered sets of four atoms that define the dihedral angle {(3126, 3127, 3128), (4126, 4127, 4128), (5126, 5127, 5128)} are calculated in the *absence* of an applied **E**-field and are used to inform the choice of dihedral angle for the **E**-field calculations. The dihedral atom numbering is provided in **Scheme 2**.

In this work we use tighter SCF convergence criteria and greater accuracy of the two-electron integrals compared to [36], which results in differences in the values of the chirality $\mathbb{C}_\sigma$, but not in the value of the chirality-helicity function $\mathbb{C}_{helicity}$ ($\approx 0$) nor the form of the variation of the ellipticity ε with torsion θ.

Single-point calculations were then undertaken on each scan geometry, where the SCF iterations were converged to < $10^{-10}$ RMS change in the density matrix and < $10^{-8}$ maximum change in the density matrix to yield the final wavefunctions for analysis. QTAIM and stress tensor analysis was performed with the AIMAll[30] and QuantVec[35] suite on each wavefunction obtained in the previous step. All molecular graphs were additionally confirmed to be free of non-nuclear attractor critical points.

## 3. Results and discussions

The scalar distance measures geometric bond-length (GBL) and bond-path length (BPL) used in this investigation are insufficient to quantify chiral effects or to understand the balance of steric effects to hyper-conjugation within the NG-QTAIM interpretation in the absence or presence of an applied ±**E**-field and are provided in the **Supplementary Materials S2**. The scalar measures for ethane with and without an applied **E**-field are provided in the **Supplementary Materials S3**.

All of the distance measures used for C-C bonds are very insensitive to the application of the ±**E**-field. Values of the distance measures of the C-H bonds increase or decrease depending on the direction of the **E**-field relative to the C-H bond, but overall do not provide any perspective as to the origins of the staggered conformation of the ethane molecule.

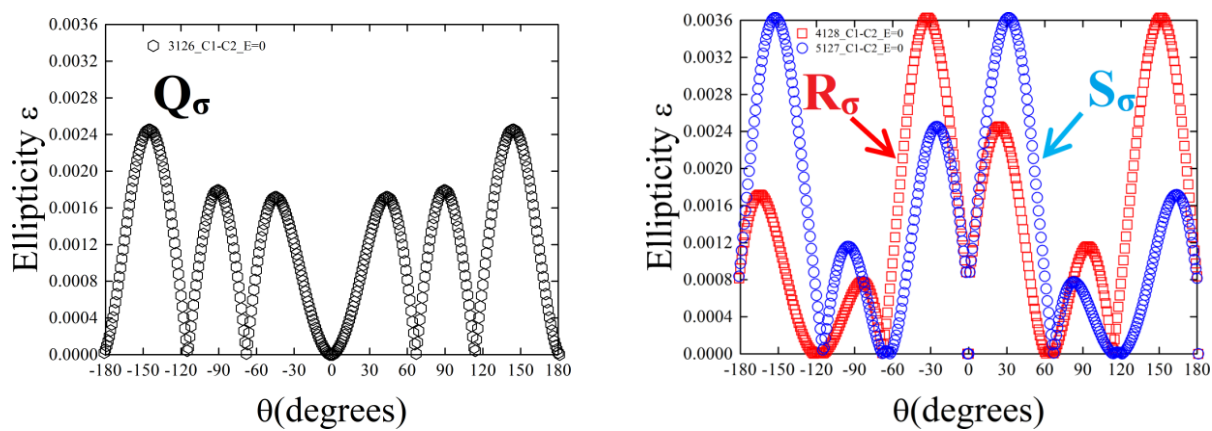

**Figure 1.** The variation of the ellipticity ε for the clockwise (CW) (-180.0° ≤ θ ≤ 0.0°) and counter-clockwise (CCW) (0.0° ≤ θ ≤ +180.0°) torsion θ isomers of the ethane C1-C2 *BCP* in the absence of an **E**-field. For the $\mathbf{Q}_\sigma$ isomer (black, left-panel) the CCW and CW portions are symmetrical about torsion θ = 0.0°. The corresponding $\mathbf{R}_\sigma$ (red) and $\mathbf{S}_\sigma$ (blue) isomers are presented in the right-panel. Note, the ellipticity ε at the torsion C1-C2 *BCP* = 0 (left and right panels).

In the current investigation a representative selection of three $\mathbb{U}_\sigma$-space distortion sets {$\mathbb{C}_\sigma,\mathbb{F}_\sigma,\mathbb{A}_\sigma$} and chirality helicity function $\mathbb{C}_{helicity} = \mathbb{C}_\sigma|\mathbb{A}_\sigma|$ of the C1-C2 *BCP* $\mathbb{T}_\sigma(s)$ of ethane for **E**-field = 0 is presented, see **Table 1** and **Figures 1-2**. The three symmetry inequivalent {$\mathbb{C}_\sigma,\mathbb{F}_\sigma,\mathbb{A}_\sigma$} are formed using the dihedral angles (3126, 4128 and 5127), which correspond to the $D_{3126}$, $D_{4128}$ and $D_{5127}$ isomers in $\mathbb{U}_\sigma$-space, see **Scheme 2**. Previously[36], we considered only one of the nine possible $\mathbb{U}_\sigma$-space distortion sets of ethane (**E**-field = 0) and discovered a $\mathbb{U}_\sigma$-space isomer with chirality assignment $\mathbf{Q}_\sigma$, (since $\mathbb{C}_{helicity} = \mathbb{C}_\sigma|\mathbb{A}_\sigma| \approx 0$). In the current investigation, the variation of the ethane C1-C2 *BCP* ellipticity ε for the CW and CCW torsions, in the absence and presence of an **E**-field, displays mirror symmetry at θ = 0.0º for the $D_{3126}$ isomer, which possesses chirality $\mathbb{C}_\sigma$ assignment $\mathbf{Q_\sigma}$, see **Figure 1**(left-panel) and the **Supplementary Materials S3**. Mirror symmetry is also present at θ = 0.0º for the ellipticity ε variations for the (CCW $D_{5127}$ and CW $D_{4128}$) and (CW $D_{5127}$ and CCW $D_{4128}$) isomers where the $D_{4128}$ and $D_{5127}$ isomers possess chirality $\mathbb{C}_\sigma$ assignments $\mathbf{R_\sigma}$ and $\mathbf{S_\sigma}$ respectively.

The remaining six $\mathbb{U}_\sigma$-space distortion sets {$\mathbb{C}_\sigma,\mathbb{F}_\sigma,\mathbb{A}_\sigma$} are provided for the absence of an applied **E**-field and the results for the symmetry inequivalent C1-C2 *BCP* and the C-H *BCP*s for the presence of the applied **E**-field are provided in the **Supplementary Materials 5**.

The individual components $\mathbb{C}_\sigma$, $\mathbb{F}_\sigma$, $\mathbb{A}_\sigma$ for the $D_{4128}$ and $D_{5127}$ isomers possess opposite assignments for the $\mathbb{T}_\sigma(s)$ i.e. {$\mathbb{C}_\sigma$ = **[$\mathbf{R_\sigma}$]**, $\mathbb{F}_\sigma$ = **[$\mathbf{R_\sigma}$]**, $\mathbb{A}_\sigma$ = **[$\mathbf{R_\sigma}$]**} and {$\mathbb{C}_\sigma$ = **[$\mathbf{S_\sigma}$]**, $\mathbb{F}_\sigma$ = **[$\mathbf{S_\sigma}$]**, $\mathbb{A}_\sigma$ = **[$\mathbf{S_\sigma}$]**}, but there are insignificant differences in the magnitudes in each of the {$\mathbb{C}_\sigma,\mathbb{F}_\sigma,\mathbb{A}_\sigma$} for the $D_{4128}$ and $D_{5127}$ isomers, see **Table 1**. Consequently, the $D_{4128}$ and $D_{5127}$ isomers can be regarded as stereoisomers in $\mathbb{U}_\sigma$-space. The magnitudes of the chirality $\mathbb{C}_\sigma$ of the $D_{4128}$ and $D_{5127}$ isomers is non-negligible (= ±0.291), particularly considering that ethane is a formally achiral molecule, however, the axiality $\mathbb{A}_\sigma$ values are very low (= ±0.001). The values of $\mathbb{C}_{helicity}$ are -0.0003 and 0.0003 for the $D_{4128}$ and $D_{5127}$ isomers, respectively, for the $\mathbb{T}_\sigma(s)$ constructed from $D_{4128}$ and $D_{5127}$. Consequently, the $\mathbb{U}_\sigma$-space isomers are again definable as stereoisomers in $\mathbb{U}_\sigma$-space, see **Table 1**. The linear sum of the $\mathbb{C}_\sigma$, $\mathbb{F}_\sigma$, $\mathbb{A}_\sigma$, $\sum\{\mathbb{C}_\sigma, \mathbb{F}_\sigma, \mathbb{A}_\sigma\}$ = 0 with corresponding chirality assignment $\mathbf{Q_\sigma}$ for the symmetry inequivalent $\mathbb{T}_\sigma(s)$ of the $D_{3126}$, $D_{4128}$, $D_{5127}$ isomers. The chirality-helicity function $\mathbb{C}_{helicity}$ ≈ 0 for the $\mathbb{T}_\sigma(s)$ of $D_{3126}$ with chirality assignment $\mathbf{Q_\sigma}$ corresponding to 'null-chirality'.

The C1-C2 *BCP* $\mathbb{T}_\sigma(s)$ for the CW and CCW torsions of the $D_{3126}$, $D_{4128}$ and $D_{5127}$ isomers overlap in a vertical plane in $\mathbb{U}_\sigma$-space, see the black spheres for the absence of an **E**-field in **Figure 2** and presence of an **E**-field in the **Supplementary Materials S5**. This finding is consistent with the mirror symmetry present in the variation of the ellipticity ε with torsion θ, see **Figure 1**. The $\mathbb{U}_\sigma$-space distortion sets {$\mathbb{C}_\sigma$, $\mathbb{F}_\sigma$, $\mathbb{A}_\sigma$} do not respond proportionally to changes in the magnitude (100×10$^{-4}$ a.u, 50×10$^{-4}$ a.u.) or (±) direction of the applied **E**-field due to the value of the torsion C1-C2 *BCP* ellipticity ε ≈ 0. The response present in the {$\mathbb{C}_\sigma$, $\mathbb{F}_\sigma$, $\mathbb{A}_\sigma$}of the C1-C2 *BCP* when the **E**-field is applied is due to the response of the C-H *BCP*s to the **E**-field affecting the C1-C2 *BCP*. The application of the **E**-field causes the value of chirality $\mathbb{C}_\sigma$ to decrease by a factor of approximately two, whereas the bond-flexing $\mathbb{F}_\sigma$ increases by more than a factor of two and an increase in the bond-axiality $\mathbb{A}_\sigma$ is noticed. The decrease in the value of chirality $\mathbb{C}_\sigma$ with the application of the **E**-field indicates a decrease in the steric effects in $\mathbb{U}_\sigma$-space, as the preference for CCW compared with

CW torsions has been reduced. The increase in $\mathbb{A}_\sigma$, although smaller than for $\mathbb{C}_\sigma$, indicates an increase in hyper-conjugation in $\mathbb{U}_\sigma$-space due to increased asymmetry in the CCW vs CW bond-path stretching.

**Table 1.** The torsion C1-C2 *BCP* $\mathbb{U}_\sigma$-space distortion sets $\{\mathbb{C}_\sigma, \mathbb{F}_\sigma, \mathbb{A}_\sigma\}$, the sum $\sum\{\mathbb{C}_\sigma, \mathbb{F}_\sigma, \mathbb{A}_\sigma\}$ and the chirality-helicity function $\mathbb{C}_{helicity}$ with the total chirality $\mathbb{C}_{\sigma T}$ and $\mathbb{A}_\sigma$ chirality assignments denoted by $[\mathbb{C}_\sigma, \mathbb{A}_\sigma]$ in the absence and presence of an applied $-100\times10^{-4}$ a.u. **E**-field. The four-digit sequence in the left column refers to the atom numbering used in the dihedral angles ($D_{3126}, D_{4128}, D_{5127}$) used to construct the stress tensor trajectories $\mathbb{T}_\sigma(s)$, which also correspond to the $D_{3126}, D_{4128}, D_{5127}$ isomer names respectively, see **Scheme 2**. The $-100\times10^{-4}$ a.u. **E**-field is directed along the C1-H3 *BCP* bond-path to create the $A_{3E\sigma}$ isomers. The corresponding ±**E**-field results for $\pm 50\times10^{-4}$ a.u. and $+100\times10^{-4}$ a.u. for the C1-H4 *BCP* ($A_{5E\sigma}$ isomers) and C1-H5 *BCP* ($A_{5E\sigma}$ isomers) are equivalent and are provided in the **Supplementary Materials S4**.

| *Isomer* | $\{\mathbb{C}_\sigma, \mathbb{F}_\sigma, \mathbb{A}_\sigma\}$ | $\mathbb{C}_{helicity}$ | $[\mathbb{C}_{\sigma T}, \mathbb{A}_\sigma]$ |
|---|---|---|---|
| **E**-field = 0 | | | |
| $D_{3126}$ | {-0.00010[$\mathbf{R}_\sigma$],0.00003[$\mathbf{S}_\sigma$],-0.000003[$\mathbf{R}_\sigma$]} | ≈ 0 (2.47×10⁻¹⁰) | [$\mathbf{Q}_\sigma$] |
| $D_{4128}$ | {-0.29110[$\mathbf{R}_\sigma$],-0.12644[$\mathbf{R}_\sigma$],-0.00094[$\mathbf{R}_\sigma$]} | -0.0003 | [$\mathbf{R}_\sigma,\mathbf{R}_\sigma$] |
| $D_{5127}$ | { 0.29096[$\mathbf{S}_\sigma$], 0.12635[$\mathbf{S}_\sigma$], 0.00099[$\mathbf{S}_\sigma$]} | 0.0003 | [$\mathbf{S}_\sigma,\mathbf{S}_\sigma$] |
| $\sum\{\mathbb{C}_\sigma, \mathbb{F}_\sigma, \mathbb{A}_\sigma\}$ | {-0.00024[$\mathbf{R}_\sigma$],-0.00006[$\mathbf{R}_\sigma$],0.000047[$\mathbf{S}_\sigma$]} | ≈ 0 (-1.13×10⁻⁸) | [$\mathbf{Q}_\sigma$] |
| **E**-field = -100×10⁻⁴ a.u | | | |
| $D_{3126}$ | {-0.00020 [$\mathbf{R}_\sigma$],-0.00011[$\mathbf{R}_\sigma$],-0.00004[$\mathbf{R}_\sigma$]} | ≈ 0 (7.06×10⁻⁹) | [$\mathbf{Q}_\sigma$] |
| $D_{4128}$ | {-0.14638[$\mathbf{R}_\sigma$],-0.28745[$\mathbf{R}_\sigma$],-0.00124[$\mathbf{R}_\sigma$]} | -0.0002 | [$\mathbf{R}_\sigma,\mathbf{R}_\sigma$] |
| $D_{5127}$ | {0.14677[$\mathbf{S}_\sigma$],0.28763[$\mathbf{S}_\sigma$],0.00112[$\mathbf{S}_\sigma$]} | 0.0002 | [$\mathbf{S}_\sigma,\mathbf{S}_\sigma$] |
| $\sum\{\mathbb{C}_\sigma, \mathbb{F}_\sigma, \mathbb{A}_\sigma\}$ | {0.000191[$\mathbf{S}_\sigma$],0.000073[$\mathbf{S}_\sigma$],-0.000158[$\mathbf{R}_\sigma$]} | ≈ 0 (-3.02×10⁻⁸) | [$\mathbf{Q}_\sigma$] |

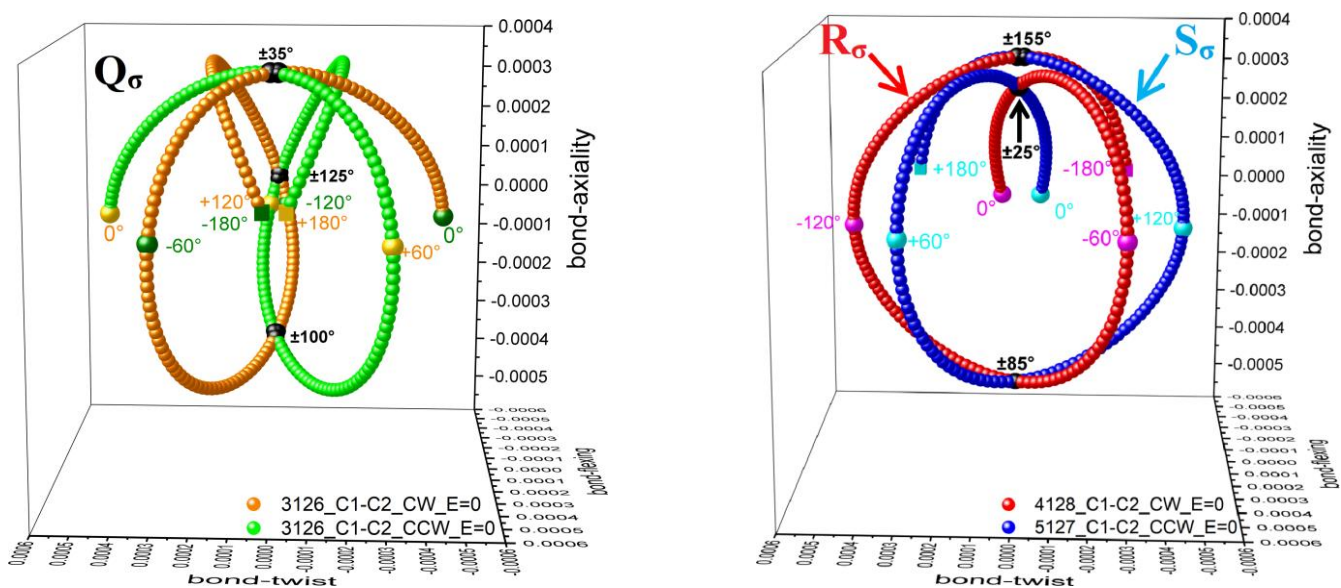

**Figure 2.** The ethane C1-C2 *BCP* stress tensor trajectories $\mathbb{T}_\sigma(s)$ in the absence of an **E**-field for the Cartesian CW and CCW torsions for the $\mathbf{Q}_\sigma$ (top-left) and $\mathbf{S}_\sigma$ (top-right, blue) and $\mathbf{R}_\sigma$ (top-right, red) $\mathbb{U}_\sigma$-space isomers. The $\mathbb{U}_\sigma$-space coordinates for the overlap (black spheres) of the CCW and CW of the $\mathbf{Q}_\sigma$ $\mathbb{T}_\sigma(s)$ are ±35°(0.0001,0.0001,0.0004), ±100°(0.0000,0.0000,-0.0003), ±125°(0.0000,-0.0003,0.0000); the corresponding locations for the $\mathbf{S}_\sigma$ and $\mathbf{R}_\sigma$ $\mathbb{U}_\sigma$-space isomers are ±25°(0.0000,0.0000,0.0003), ±85°(0.0000,0.0002,-0.0003), ±155°(-0.0001,-0.0001,0.0003).

**Conclusions**

In this investigation the resultant chiral nature of the ethane molecule was determined to be achiral in $\mathbb{U}_\sigma$-space. Due to the vector-based nature of $\mathbb{U}_\sigma$-space it was possible to obtain this resultant chiral character from additive contributions. This result was found using a newly developed extension of our NG-QTAIM chirality analysis that searched all nine possible torsion C1-C2 *BCP* stress tensor trajectories $\mathbb{T}_\sigma(s)$ to determine the three symmetry inequivalent $\mathbb{T}_\sigma(s)$. The finding that ethane is achiral in $\mathbb{U}_\sigma$-space is consistent with the result of summing the values of the chirality helicity function $\mathbb{C}_{helicity}$ ($\approx 0$) for the $D_{3126}$, $D_{4128}$ and $D_{5127}$ isomers. Previously, our NG-QTAIM analysis was limited to using only one $\mathbb{T}_\sigma(s)$ for each molecular graph to determine the $\mathbf{R}_\sigma$ or $\mathbf{S}_\sigma$ chirality $\mathbb{C}_\sigma$ assignments.

In this work the resultant chirality $\mathbb{C}_\sigma$ of ethane was determined to be the newly introduced chirality assignment $\mathbf{Q}_\sigma$, which can be regarded as a 'null-chirality'. The resultant $\mathbf{Q}_\sigma$ chiral character of ethane was determined by summing the individual $\mathbb{C}_\sigma$, $\mathbb{F}_\sigma$ and $\mathbb{A}_\sigma$ components of the $\mathbb{U}_\sigma$-space distortion sets {$\mathbb{C}_\sigma$, $\mathbb{F}_\sigma$, $\mathbb{A}_\sigma$} of the $D_{3126}$, $D_{4128}$ and $D_{5127}$ isomers of the torsion C1-C2 *BCP*. The $D_{4128}$ and $D_{5127}$ isomers were determined to be stereoisomers in $\mathbb{U}_\sigma$-space in the absence and presence of an applied **E**-field. The implicit presence of large chirality $\mathbb{C}_\sigma$ values for the $D_{4128}$ and $D_{5127}$ isomers of ethane indicates relevance for understanding why steric effects are a significant factor in explaining the staggered geometry of ethane. This is because significant values of the chirality $\mathbb{C}_\sigma$ indicate an asymmetry in the CCW vs CW torsion for the C1-C2 *BCP*, i.e. relating to steric effects in $\mathbb{U}_\sigma$-space. These chirality $\mathbb{C}_\sigma$ values for the $D_{4128}$ and $D_{5127}$ isomers are also consistent with equal and opposite torsions of the $CH_3$ groups located in a staggered configuration either side of the torsional C1-C2 *BCP* bond-path of the relaxed structure of ethane. The dominance of the chirality $\mathbb{C}_\sigma$ values of the $D_{4128}$ and $D_{5127}$ isomers, which are two orders of magnitude greater than the corresponding values of the C1-C2 *BCP* bond-axiality $\mathbb{A}_\sigma$, indicate the dominance of steric effects over hyper-conjugation within the NG-QTAIM interpretation. The lack of significant values of the C1-C2 *BCP* bond-axiality $\mathbb{A}_\sigma$ indicates a lack of asymmetry in the response of the torsion C1-C2 *BCP* bond-path to the CCW vs CW torsion. The steric effects were reduced by a factor of two with the application of the **E**-field, although no measurable changes in the ethane molecular geometry were observed since the steric effects are determined in $\mathbb{U}_\sigma$-space.

The $\mathbb{U}_\sigma$-space distortion sets {$\mathbb{C}_\sigma$, $\mathbb{F}_\sigma$, $\mathbb{A}_\sigma$} of the torsion C1-C2 *BCP* did not respond proportionally to the applied **E**-field, i.e. the response was independent of the direction (+) or (-) or the magnitude of the applied **E**-field due to value of the torsion C1-C2 *BCP* ellipticity $\varepsilon = 0$ for the relaxed ethane molecular graph. The applied **E**-field produced indirect effects on the C1-C2 *BCP* $\mathbb{U}_\sigma$-space distortion sets {$\mathbb{C}_\sigma$, $\mathbb{F}_\sigma$, $\mathbb{A}_\sigma$} compared to the absence of the applied **E**-field via the response of the C-H *BCP*s.

Future investigations of chiral and formally achiral molecules could be undertaken using $\mathbb{T}_\sigma(s)$ constructed from a summation of contributions from the symmetry inequivalent set of $\{\mathbb{C}_\sigma, \mathbb{F}_\sigma, \mathbb{A}_\sigma\}$ of the $\mathbb{T}_\sigma(s)$ of the torsion *BCP* associated with any suspected chiral center, instead of using a single $\{\mathbb{C}_\sigma, \mathbb{F}_\sigma, \mathbb{A}_\sigma\}$ as we previously undertook. For achiral and chiral reactions the symmetry inequivalent set of isomers with $\mathbf{Q}_\sigma$, $\mathbf{S}_\sigma$ or $\mathbf{R}_\sigma$ chirality assignments could be kept separate and tracked for the duration of the reaction. The symmetry inequivalent set of isomers with $\mathbf{Q}_\sigma$, $\mathbf{S}_\sigma$ or $\mathbf{R}_\sigma$ chirality assignments would be chosen from the reactant to transition state portion of the reaction pathway. The corresponding set of isomers with $\mathbf{Q}_\sigma$, $\mathbf{S}_\sigma$ or $\mathbf{R}_\sigma$ chirality assignments would be chosen from the product.

**Funding Information**

The National Natural Science Foundation of China is gratefully acknowledged, project approval number: 21673071. The One Hundred Talents Foundation of Hunan Province is also gratefully acknowledged for the support of S.J. and S.R.K. H.F. and T.v.M. gratefully acknowledge computational support via the EaStCHEM Research Computing Facility.